\documentclass[twocolumn,showpacs,superscriptaddress,preprintnumbers,amsmath,amssymb]{revtex4-2}
\usepackage{amssymb}		
\usepackage{amsmath}
\usepackage{graphicx}
\usepackage[normalem]{ulem}
\usepackage{multirow}
\usepackage{appendix}
\usepackage{CJK}
\usepackage[usenames]{color}
\usepackage{bm}
\usepackage[colorlinks,linkcolor=blue,urlcolor=blue,citecolor=blue]{hyperref}
\usepackage{booktabs} 	
\usepackage{leftidx}

\usepackage{ulem}
\newcommand\redsout{\bgroup\markoverwith{\textcolor{red}{\rule[0.5ex]{2pt}{0.4pt}}}\ULon}

\usepackage[all]{hypcap} 
\makeatletter

\newcommand{\Rmnum}[1]{\expandafter\@slowromancap\romannumeral #1@}
\makeatother

\usepackage{tikz,xcolor,hyperref}

\definecolor{lime}{HTML}{A6CE39}
\DeclareRobustCommand{\orcidicon}{
	\begin{tikzpicture}
	\draw[lime, fill=lime] (0,0) 
	circle [radius=0.16] 
	node[white] {{\fontfamily{qag}\selectfont \tiny ID}};
	\draw[white, fill=white] (-0.0625,0.095) 
	circle [radius=0.007];
	\end{tikzpicture}
	\hspace{-2mm}
}
\foreach \x in {A, ..., Z}{%
	\expandafter\xdef\csname orcid\x\endcsname{\noexpand\href{https://orcid.org/\csname orcidauthor\x\endcsname}{\noexpand\orcidicon}}
}

\begin{document}
	\begin{CJK*} {UTF8} {gbsn}

\title{Specific heat and its high-order moments in relativistic heavy-ion collisions from a multiphase transport model}

\author{Ru-Xin Cao(曹汝鑫)}

\affiliation{Shanghai Institute of Applied Physics, Chinese Academy of Sciences, Shanghai 201800, China}
\affiliation{Key Laboratory of Nuclear Physics and Ion-beam Application (MOE), Institute of Modern Physics, Fudan University, Shanghai 200433, China}
\affiliation{School of Nuclear Sciences and Technology, University of Chinese Academy of Sciences, Beijing 100049, China}

\author{Song Zhang(张松)\orcidB{} 
}
\thanks{Email: song\_zhang@fudan.edu.cn}
\affiliation{Key Laboratory of Nuclear Physics and Ion-beam Application (MOE), Institute of Modern Physics, Fudan University, Shanghai 200433, China}
\affiliation{Shanghai Research Center for Theoretical Nuclear Physics， NSFC and Fudan University, Shanghai 200438, China}

\author{Yu-Gang Ma(马余刚)\orcidC{}}\thanks{Email:  mayugang@fudan.edu.cn}
\affiliation{Key Laboratory of Nuclear Physics and Ion-beam Application (MOE), Institute of Modern Physics, Fudan University, Shanghai 200433, China}
\affiliation{Shanghai Research Center for Theoretical Nuclear Physics， NSFC and Fudan University, Shanghai 200438, China}
\date{\today}

\begin{abstract}
Energy dependence of specific heat extracted from  temperature fluctuation
of Au + Au collisions at $\sqrt{s_{NN}}$ =  7.7 GeV to 200 GeV was investigated by using a multiphase transport  (AMPT) model.  The results were compared with those from other models and  some differences at low $\sqrt{s_{NN}}$ were found. To explain the above differences and describe the properties of the hot dense matter at low $\sqrt{s_{NN}}$, a new quantity 
$C_v^{*}$  was derived for describing  specific heat in heavy-ion collisions. It was found that, by using $C_v^{*}$ together with its high order moments (skewness and kurtosis), thermal properties of the hot dense matter can be  described and different thermal properties with or without parton process  can be clearly distinguished. The proposed observable provides a way to learn  the property of QCD matter 
in heavy-ion collisions.
\end{abstract}
\maketitle

\section{Introduction}

 During the past few decades, amount efforts have been made on studies of the hot dense quark matter created in relativistic heavy-ion collisions. Plenty of evidences supported the existence of quark-gluon plasma (QGP) in relativistic heavy-ion collisions and aroused the interest to explore the properties of hot dense matter created under the extreme conditions  of temperature and density \cite{Shuyak,Bzdak,PBM,Chen,Tang_NST,Shen_NST,Mrow,Luo2017,Song_CPL,Liu}. Specific heat was carried out as one of the signals of phase transition in partonic level as well as nucleonic level and helps to inspect thermal properties of nuclear matter. 
 In partonic level, the Lattice QCD predicted the phase transition as a crossover at zero baryon-chemical potential $(\mu_{B}=0)$ but while a first order phase transition could occur at finite baryon density
 ~\cite{temperature_fluctuation_thermal_A,temperature_fluctuation_Ayala1,temperature_fluctuation_Ayala2,temperature_fluctuation_Basu,temperature_fluctuation_Cv_Shuryak,temperature_fluctuation_Stodolsky,Han}. In nucleonic level, a liquid gas phase transition can occur at sub-saturation density and moderate temperature \cite{Pan,YGMa1999,Ago,JBNatowitz2002,Chomaz,Ma2005,CWMa2018,Lin2019,BBorderie2019,Liuhuanling,Deng22,LiuC_NST}. 

According to phase transition theory, 
 the long-range correlation diverges rapidly when a thermodynamic system evolves close to critical point. Here it is relevant to the  degree of freedom of  the quarks and gluons, rather than mesons and baryons for the matter created in the collisions. As a response to the system perturbation the specific heat  could diverge when the system evolves close to the critical point. In statistical physics the heat capacity can be associated with the fluctuation of temperature.
And many works tried to  extract heat capacity from the event-by-event distribution of ensemble temperature~\cite{temperature_fluctuation_Stodolsky,MaGL04,Ma05,Shi07,Cv_pT_HM_QGM_Sa,Cv_models_mean_pT_Basu,event_by_event_fluctuation_Jeon,temperature_fluctuation_Ayala1,temperature_fluctuation_Ayala2,temperature_fluctuation_Basu,temperature_fluctuation_Cv_Shuryak,temperature_fluctuation_thermal_A,Ma2005,Lin2019}. At near the critical point, the specific heat $C_V$ can be expressed as a power law function of  temperature deviated from critical temperature ($T_c$), i.e.  $C_V \propto |T-T_c|^{-\alpha}$ 
with $\alpha$ as a critical exponent. Furthermore, although energy dependence of the specific heat $C_V$ has been  studied in some theoretical works~\cite{Cv_pT_HM_QGM_Sa,Cv_models_mean_pT_Basu,Cv_mean_pT_Pythia_ALICE_Deb}, event-by-event fluctuation of the specific heat itself should be also paid more attention for investigating its statistical properties as a potential probe to the QCD critical point.

To this end, a multiphase transport (AMPT) model~\cite{AMPT2005,AMPTGLM2016,AMPT2021} was adopted to simulate Au + Au collisions at various energies of $\sqrt{s_{NN}}$ = 7.7, 11.5, 14.5, 19.6, 27, 39, 62.4, and 200 GeV. Energy dependence of $C_V$ was obtained and compared with results extracted through the same temperature fluctuation method by other models, namely the Hadron Matter (HM)~\cite{Cv_pT_HM_QGM_Sa}, Hadron Matter via Quark Gluon Matter~\cite{Cv_pT_HM_QGM_Sa}, Quark Gluon Matter (QGM)~\cite{Cv_pT_HM_QGM_Sa}, Hadron Resonance Gas (HRG) models as well as  the data deduced from the STAR Collaboration \cite{Cv_models_mean_pT_Basu,Cv_mean_pT_Pythia_ALICE_Deb}.  However, it was found that the specific heat obtained from the AMPT model shows  obvious depression at low $\sqrt{s_{NN}}$, similar to some other models~\cite{Cv_models_mean_pT_Basu,Cv_mean_pT_Pythia_ALICE_Deb,Cv_pT_HM_QGM_Sa}. This insensitivity phenomenon of temperature fluctuation in the AMPT model drove our present study to propose another derivation of heat capacity from the basic definition with some assumptions. The derivation in this study gives  an effective specific heat, denoted as $C_v^{*}$, which is expressed by the characteristic event's  
kinetic quantities, and plots the event-by-event distribution to obtain the skewness and kurtosis to describe the statistical properties of $C_v^{*}$. The mean value of new derived $C_v^{*}$ together with its skewness and kurtosis give insight of the evolution of thermodynamical properties of the emitted particles and describe energy dependence of specific heat, which could be performed in experiments.

The paper is organized as follows: in Sec.~\ref{sec:AMPT},  a brief introduction of a multiphase transport  model was given, then the results on the energy dependence of specific heat from the event-by-event temperature fluctuation in Au + Au collisions simulated by the AMPT were present, and  the previous results from other models as well as the STAR data were compared. By the above comparison, the difference and the uncertainty of specific heat  at low $\sqrt{s_{NN}}$ in our AMPT calculation were analyzed. 
In Sec.~\ref{sec:derivation}, a brief qualitative explanation for the differences was presented, and   a formula expressed by kinetic quantities called $C_v^{^*}$ from the definition of specific heat was proposed. 
Sec.~\ref{sec:results} gave the results for mean value, skewness, kurtosis of event-by-event distributions for $C_v^{*}$, from which 
 the meaning of those moments was explained 
 by emitted particles' thermodynamic properties. Finally a brief summary was given in Sec.~\ref{sec:summary}.

\section{AMPT results and comparison}
\label{sec:AMPT}
\subsection{Introduction to AMPT}

A multiphase transport (AMPT) model~\cite{AMPT2005,AMPTGLM2016,AMPT2021} is composed of four stages to simulate relativistic heavy-ion collisions. It has successfully described various phenomena at RHIC and LHC energies and becomes a well-known event generator. The AMPT has two versions: String Melting (SM) and Default. In SM version, Heavy Ion Jet Interaction Generator (HIJING)~\cite{HIJING-1,HIJING-2} is used to simulate the initial conditions, then Zhang's Parton Cascade (ZPC)~\cite{ZPCModel} is used to describe interactions for partons which are from all of hadrons in the HIJING but spectators, after which a simple Quark Coalescence Model describes hadronization process, finally A Relativistic Transport (ART) model~\cite{ARTModel}  simulates hadron re-scattering process. The Default version of AMPT only conducts the minijet partons in partonic scatterings via ZPC and uses the Lund string fragmentation to perform hadronization.

AMPT model  ~\cite{AMPT2005,AMPT2021} can describe the $p_T$ spectrum and energy dependence of identified particles such as pion, kaon, $\phi$, proton and $\Omega$ produced in heavy-ion collisions ~\cite{AMPTGLM2016,Jin2018,Wang2019},  as well as the collective flows and temperature during evolution etc~\cite{AMPT_temperature_parton_Lin,HRG_model_Mait,Cv_models_mean_pT_Basu,NSTSongFlow}. Chiral and magnetic related anomalous phenomena can also be described by the AMPT model \cite{ZhaoXL2019,Gao2020,Huang2020,WangCZ2021,Zhao2019,WuWY}. More details of the model description and selection of the set for parameters can be found in Refs.~\cite{AMPT2005,AMPTGLM2016,AMPT2021}. The  present study used both versions to simulate Au + Au collisions at $\sqrt{s_{NN}}$ = 7.7, 11.5, 14.5, 19.6, 27, 39, 62.4, and  200 GeV in the impact parameter range of 0 - 4.7 fm which is corresponding to centrality 0 - 10\%.

\subsection{$C_V$ from temperature fluctuation}

Firstly, an event-by-event fluctuation method was used to study 
the specific heat. As noted in Refs.~\cite{Cv_models_mean_pT_Basu,Cv_pT_HM_QGM_Sa}, since the lower multiplicity of particles caused relatively large error in fitting $p_T$ spectrum,  we then merged some events into one new event, denoted as ``linked event".  The combined events' number is determined by ensuring the multiplicity in one ``linked event" more than 1000, thus we increased mean multiplicity in one ``event" more than 1000, as other researchers did in Ref.~\cite{Cv_models_mean_pT_Basu}. And as shown in Fig.~\ref{fig:skewness} of Ref.~\cite{Cv_models_mean_pT_Basu}, by comparing the results in ``linked events" and in ``random mixed events" in which each particle from different events (plotted by blue and red line), we can see ``linking  events" from different original event sets  only modify fluctuation results slightly.

The effective temperature, $T_{eff}$, of a $\pi^+$ system at final state is obtained via fitting $p_T$ spectrum by using the exponential distribution~\cite{AMPTGLM2016,pT_spectrum_mean_pT_Tkin_beta_Adam,pT_spectrum_temperature_Waqas}:
\begin{eqnarray}
\begin{aligned}
\frac{1}{p_T}\frac{dN}{dp_T} = A e^{\frac{-p_T}{T_{eff}}}.
\end{aligned}
\label{eq:Teff}
\end{eqnarray}
$\pi^+$ system is chosen for the calculations with kinetic window, $p_T < 2$ GeV/$c$ and $-1 < y < 1$.
$T_{eff}$  stands for the slope of $p_T$ spectrum and consists of two parts， i.e.  the kinetic temperature $T_{kin}$ and contribution from radial flow $\langle \beta _T\rangle$~\cite{Cv_models_mean_pT_Basu},
\begin{eqnarray}
T_{eff} = T_{kin} + f(\langle \beta _T\rangle), 
\label{Teff_Tkin_fbeta}
\end{eqnarray}
where the kinetic freeze-out temperature, $T_{kin}$, characterizes thermal motion of emitted particles, and $f(\langle \beta _T\rangle)$ reflects transverse radial flow contribution. As we are  discussing $\pi^+$ system, the relation in Eq.~(\ref{Teff_Tkin_fbeta}) can be approximately written as ~\cite{Cv_models_mean_pT_Basu},
\begin{eqnarray}
T_{eff}\approx T_{kin} + m_0\langle \beta _T\rangle^2,
\label{Teff_Tkin_m0}
\end{eqnarray}
where $m_0$ is the mass of $\pi^+$.

For a system in equilibrium, its heat capacity ($C$) is related to the fluctuation of event-by-event  temperature  distribution~\cite{event_by_event_fluctuation_Jeon,temperature_fluctuation_Basu,temperature_fluctuation_thermal_A,temperature_fluctuation_Cv_Shuryak}, 
\begin{eqnarray}
\begin{aligned}
P(T) \propto e^{-\frac{C}{2}\frac{(\Delta T)^2}{\langle T\rangle ^2}}, 
\end{aligned}
\end{eqnarray}
where $\langle T\rangle$ is the mean temperature  value 
and $\Delta T = T - \langle T\rangle$ is the deviation of temperature from its mean value. The expression can be further derived as ~\cite{Cv_models_mean_pT_Basu}:
\begin{eqnarray}
\label{eq:C_fluctuation}
\begin{aligned}
\frac{1}{C} = \frac{\langle T_{kin}^2\rangle -\langle T_{kin}\rangle ^2}{\langle T_{kin}\rangle ^2}\approx \frac{\langle T_{eff}^2\rangle -\langle T_{eff}\rangle ^2}{\langle T_{kin}\rangle ^2}.
\end{aligned}
\end{eqnarray}
When  multiplicity $N$ is taken into account, we can get specific heat per particle ($C_V$) as the following
\begin{eqnarray}
\label{eq:Cv}
\begin{aligned}
C_V = \frac{C}{N}.
\end{aligned}
\end{eqnarray}

\begin{figure}[htbp]
	\centering
	\includegraphics[angle=0,scale=0.45]{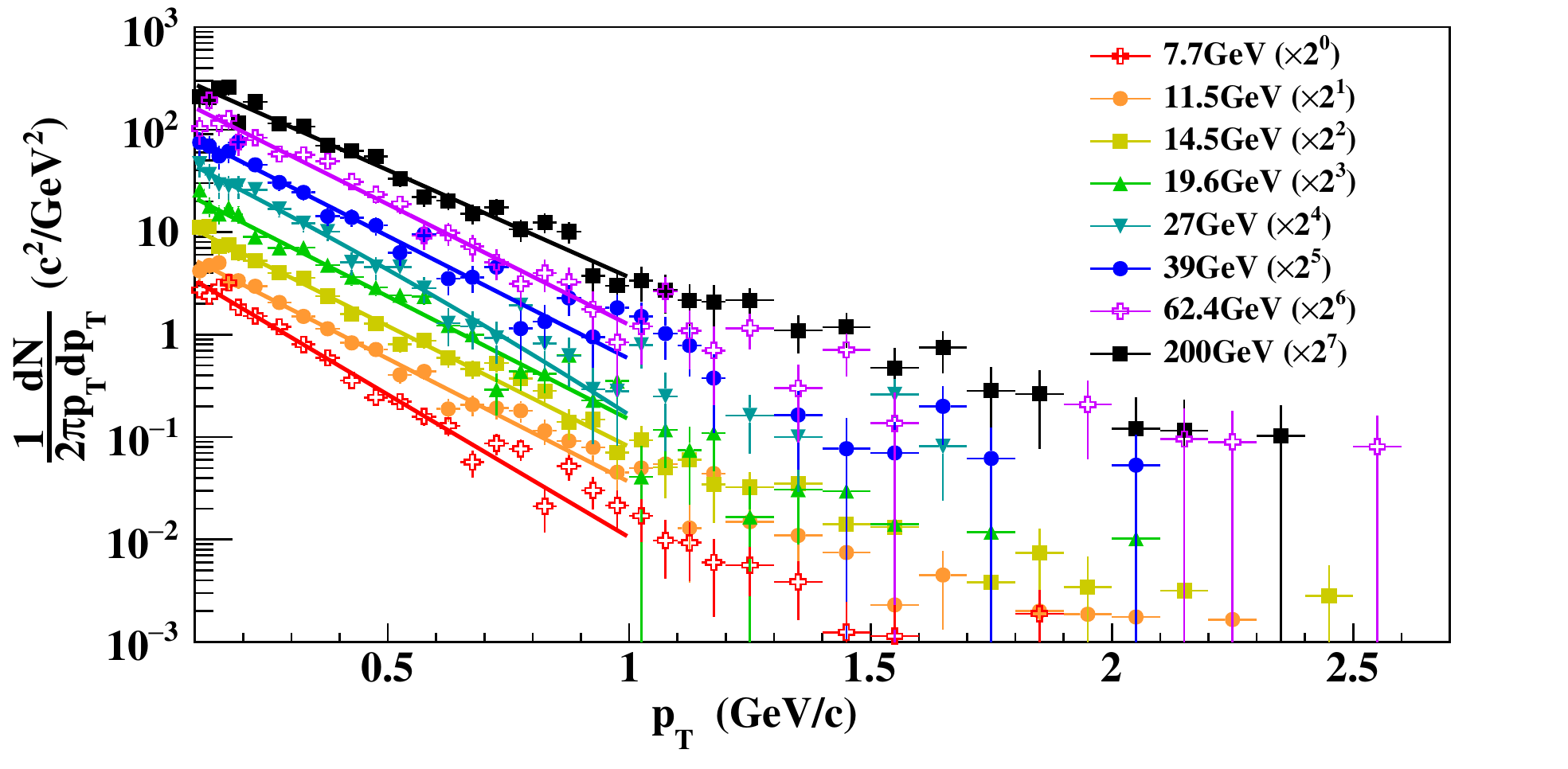}
	\caption{The $p_T$ spectrum for the merged events with $-1<y<1$ at $\sqrt{s_{NN}} = 7.7  -200$ GeV in $0-10\%$ central Au + Au collisions by using the String Melting AMPT. The line represents the Boltzmann fit to the $p_T$ spectra below 1 GeV/c.}
\label{fig:pt_spectrum}
\end{figure}

\begin{figure}[htbp]
	\centering
	\includegraphics[angle=0,width=9cm,height=9cm]{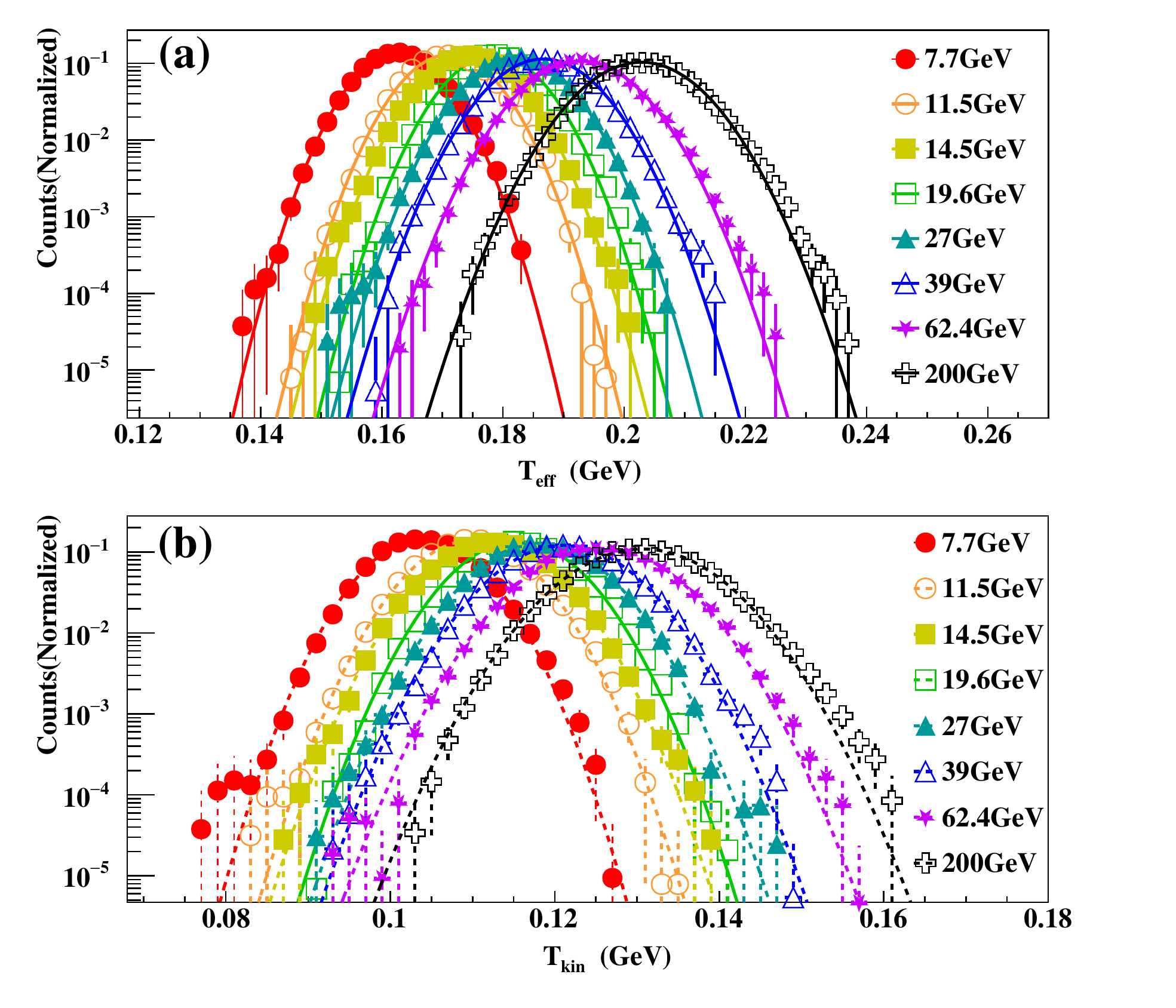}
	\caption{The  event-by-event  distributions of effective temperature (a) and kinetic temperature (b) at $\sqrt{s_{NN}} = 7.7 - 200$ GeV in $0-10\%$ central Au + Au collisions by using the String Melting AMPT model, the lines are fitted by Gaussian distribution.}  
	\label{fig:T_eff}
\end{figure}

In this section we choose $T_{kin}$ to extract $C_V$ since the variances of $T_{kin}$ and $T_{eff}$ are close enough to each other as mentioned in Ref.~\cite{Cv_models_mean_pT_Basu}.  Figure~\ref{fig:pt_spectrum} shows the $p_T$ spectra in ``linked events" fitted by Eq.~(\ref{eq:Teff}), from which the effective temperature was obtained as shown in Fig.~\ref{fig:T_eff}(a). Meanwhile, in the present work we calculated transverse flow velocity $\beta _T$ by $\beta_T = \frac{ \boldsymbol{\beta}_{\boldsymbol{p_T}}\cdot \boldsymbol{\rho}}{|\boldsymbol{\rho}|}$, where $\boldsymbol{\beta}_{\boldsymbol{p_T}}$ is the transverse velocity and $\boldsymbol{\rho}$ is the transverse coordinates of particles at kinetic freeze-out state. In Refs. ~\cite{Cv_models_mean_pT_Basu,Blast_Wave_pT_Retiere} where a Blast-Wave Model was used to obtain a fit parameter also denoted by $\beta_T$. The two definitions give the same dimension of $\beta _T$ and are both on a super-surface where particles reach kinetic freeze-out status. So the two $\beta _T$ are conceptually consistent. As mentioned in Ref.~\cite{Cv_models_mean_pT_Basu}, fluctuation of $\beta _T$ could be dominant for small, asymmetric and non-central collisions. However, our work focus on Au + Au system, the fluctuation from $\beta_T$ is almost negligible.

As we mentioned before, we used the Boltzmann distribution to extract effective temperature from $p_T$ spectra as shown in  Fig.~\ref{fig:pt_spectrum}. However, it is noticed  that as $p_T$ increases, the Boltzmann distribution becomes poor to fit the spectra. To ensure our fitting results comparable and verifiable to results in Refs.~\cite{Cv_models_mean_pT_Basu,Cv_pT_HM_QGM_Sa}, however, we limit the fitting range of $p_T$ below 1 GeV/$c$ with the Boltzmann distribution, which was represented by the lines in Fig.~\ref{fig:pt_spectrum}. Of course, some checks have been done. For examples, one way is extending the $p_T$ fit range to 3 GeV/c and another way is just checking the non-thermal tail of high $p_T$ from 1 GeV/c to 3 GeV/c with the same Boltzmann distribution. From the above fit procedures, we can roughly get the lower ($p_T <$ 1 GeV/c) and upper (1 GeV/c $< p_T <$ 3 GeV/c) limits of the event-by-event temperature distribution. Further we can get the lower and upper bounds for heat capacity. It was found that the different $p_T$ range does not change the energy dependence of $C_V$ but do induce an uncertainty  of about 30\% of the $C_V$ values.  

On the other hand, there exists a few alternative  models to fit $p_T$ spectra, eg. with the Tsallis-Blast-Wave model to describe $p_T$ spectra in a wide $p_T$ range up to 3 GeV/$c$ shown in Ref.~\cite{TBW_fitting,Su}. The Tsallis distribution introduced a new parameter $q$ to represent how the system deviates from the equilibrium state. And it can be seen in Ref.~\cite{TBW_fitting} that in our range for $p_T<$1 GeV/$c$, $q-1$ is very  close to zero. The fact that $q-1$ tends to $0$ means the Tsallis distribution degrades into the Boltzmann distribution, indicating that our fits using the  Boltzmann distribution is reasonable. Further, in Ref.~\cite{Tsallis_Cv}, the Tsallis distribution was used to investigate thermal parameters in small system ($p + p$ at $\sqrt{s_{NN}} = 7$ TeV) and fitted well in the  PYTHIA frame. Of course, adopting a new Tsallis-Blast-Wave fit to deduce the heat capacity is naturally an interesting topic, 
which deserves further investigation in future.

After removing transverse radial flow contribution from Eq.~(\ref{Teff_Tkin_m0}), the distribution of event-by-event $T_{kin}$ is presented in Fig.~\ref{fig:T_eff}(b) at the centrality of $0-10\%$ and $\sqrt{s_{NN}}$ = 7.7 - 200 GeV. 
It is clearly seen that $T_{kin}$ is much lower than $T_{eff}$ because the transverse radial flow is taken off. 
Via fitting the $T_{kin}$ distribution by Eq.~(\ref{eq:C_fluctuation}) the heat capacity per particle $C_V$ can be extracted.

\begin{figure}[htbp]
	\centering
\includegraphics[angle=0,scale=0.48]{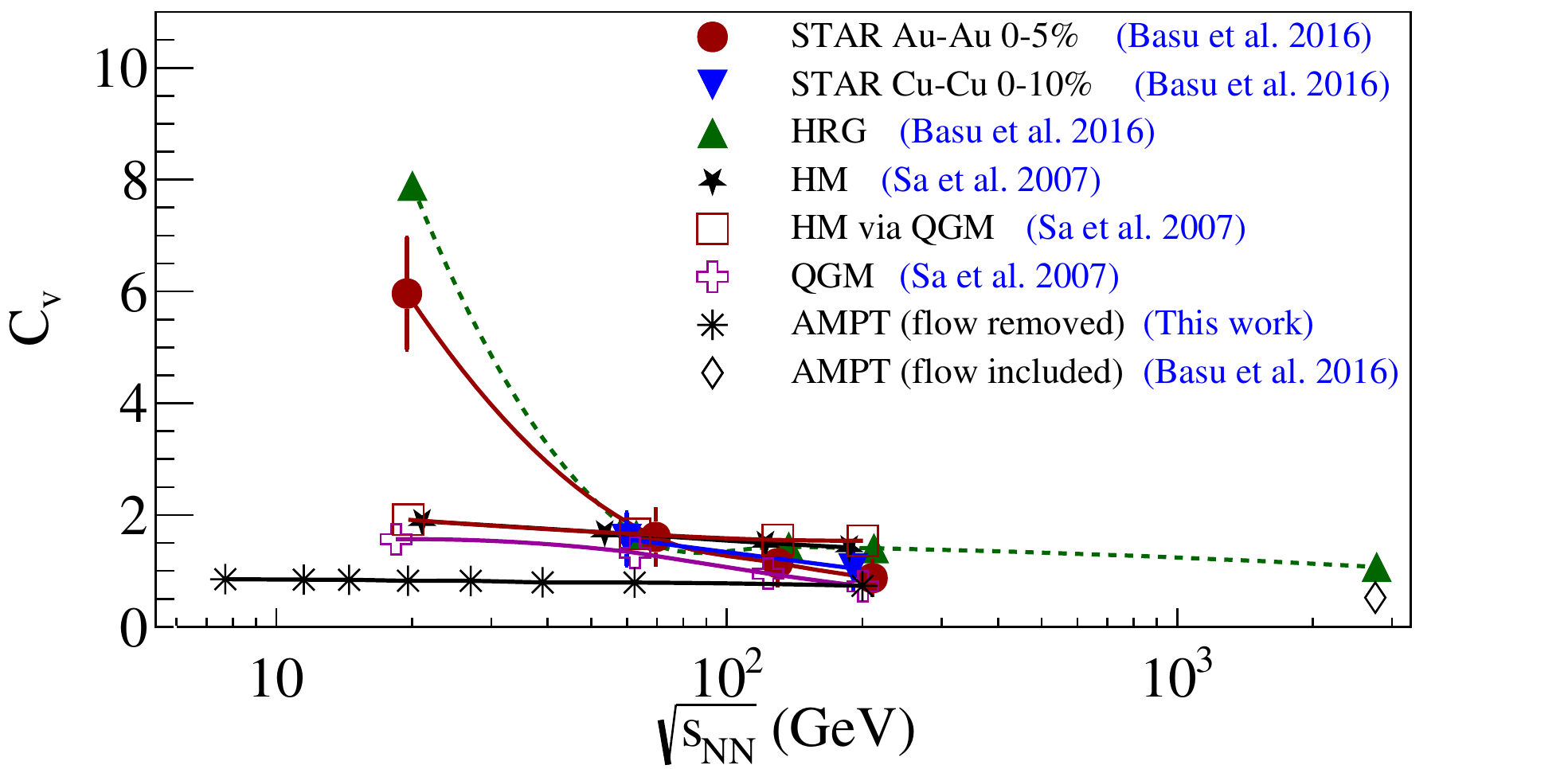}
\caption{The energy dependence of specific heat $C_v$ = $\frac{C}{N}$ is calculated via the temperature fluctuation method:  the AMPT model with removing radial flow effect (this work), the results using the STAR 0-5\% centrality Au-Au data (Basu {\it et al.} 2016 \cite{Cv_models_mean_pT_Basu}), the HRG model  (Basu {\it et al.} 2016 \cite{Cv_models_mean_pT_Basu}), the AMPT model with flow included (Basu {\it et al.} 2016~\cite{Cv_models_mean_pT_Basu}), the HM model (Sa {\it et al.} 2007 ~\cite{Cv_pT_HM_QGM_Sa}), the HM via the QGM model  (Sa {\it et al.} 2007 ~\cite{Cv_pT_HM_QGM_Sa}), and  the QGM model  (Sa {\it et al.} 2007 ~\cite{Cv_pT_HM_QGM_Sa})  .}
	\label{fig:Cv_models}
\end{figure}

Fig.~\ref{fig:Cv_models} shows different model results together with the STAR data by temperature fluctuation method. The extracted $C_V$ from the STAR data decreases with the increasing of $\sqrt{s_{NN}}$~\cite{Cv_models_mean_pT_Basu}， and it shows a significant rising trend  when $\sqrt{s_{NN}}$ goes down to $40$ GeV and below, which indicates that the system might enter critical region. The HRG model~\cite{Cv_models_mean_pT_Basu} shows  similar energy dependence of $C_V$ to the STAR Au + Au results. But other results from the HM~\cite{Cv_pT_HM_QGM_Sa}, the QGM~\cite{Cv_pT_HM_QGM_Sa}, {the AMPT ~\cite{Cv_models_mean_pT_Basu}, as well as this work employing AMPT with removing the radial flow effect give an almost constant value of $C_V$ in a wide range of $\sqrt{s_{NN}}$.} These results demonstrated a  significant model dependence at low $\sqrt{s_{NN}}$ but are consistent with each other at high $\sqrt{s_{NN}}$~\cite{Cv_models_mean_pT_Basu,Cv_mean_pT_Pythia_ALICE_Deb,Cv_pT_HM_QGM_Sa}. 

To understand the difference between  
the AMPT simulation results and the enhancement structure from the STAR data  or the HRG model~\cite{Cv_models_mean_pT_Basu} at low $\sqrt{s_{NN}}$,   
a few issues may be related as indicated in  Refs.~\cite{Cv_models_mean_pT_Basu,Cv_pT_HM_QGM_Sa}:

 (1) The finite particle multiplicity, choice of kinetic window of $p_T$ and fit parameters in Eq.~(\ref{eq:Teff}) provide uncertainties of $C_{v}$, which was mentioned in Refs.~\cite{Cv_models_mean_pT_Basu,Cv_pT_HM_QGM_Sa} as well. To check our fitting results on $p_T$ spectrum and ensure our fitting results comparable to those in Ref.~\cite{Cv_models_mean_pT_Basu}, we chose the same $p_T$ window as Ref.~\cite{Cv_models_mean_pT_Basu}.

(2) In many theoretical derivations, the system evolves in perfect thermodynamic conditions, which means that the system volume is fixed and multiplicity should be high enough. However, these conditions can hardly be satisfied in real heavy-ion collision experiments, as noted in Ref.~\cite{Cv_pT_HM_QGM_Sa};

Two reasons mentioned above are both independent of models. Besides, the model dependence should also be commented, i.e.,  different model frames provide respective additional uncertainty, which could be caused by different mechanisms used in models, even the results are calculated in the same kinetic window of transverse momentum and rapidity. 

In the following, we  try to discuss the insensitivity of temperature fluctuation in the present AMPT framework. 

\section{Derivation of a new formula of heat capacity: $C_v^{*}$}
\label{sec:derivation}

 To avoid the model dependence of extracted specific heat, the definition of heat capacity according to Ref.~\cite{Statistical_Physics_Landau} can be written as
\begin{eqnarray}
\label{eq:Cv_def}
\begin{aligned}
C = \left(\frac{\partial E_{th}}{\partial T}\right)_V,
\end{aligned}
\end{eqnarray} 
where $E_{th}$ is the energy of thermal motion in a researched system and $T$ is the temperature with a fixed volume of the system. Actually the definition of the heat capacity by Eq.~(\ref{eq:Cv_def}) and Eq.~(\ref{eq:C_fluctuation}) are equivalent. In Ref.~\cite{Statistical_Physics_Landau} there is nothing different but the one in Ref.~\ref{eq:Cv_def} is defined from differential of enthalpy in Eq. (14.6) of  Chapter 15 and the one in Ref.~\ref{eq:C_fluctuation} is derived from fluctuation in Eq. (111.6) of Chapter 12.

The definition in Eq.~(\ref{eq:Cv_def}) includes following assumptions:

(1) The energy of $\pi^+$ system at final state -- $E$ -- has positive correlation to $\sqrt{s_{NN}}$, which means for each collision at final state, total energy of $\pi^+$ system evolves along a certain curve with $\sqrt{s_{NN}}$~\cite{ET_Adams};

(2) Estimated phase volume can be approximately calculated as charged particle multiplicity $N$, which means that we can calculate specific heat per particle $C_v = \frac{C}{N}$ and the merged events are  ensured to have enough multiplicity (here over one thousand)~\cite{Cv_models_mean_pT_Basu,Cv_pT_HM_QGM_Sa};

(3) The $\pi^+$ system evolves to the kinetic freeze-out stage after the hadronic interaction. That means the the energy contributed by system's internal interaction (mainly the hadronic interaction) can be ignored when we calculated the heat capacity.

\begin{figure}[htbp]
	\centering
	\includegraphics[angle=0,scale=0.4]{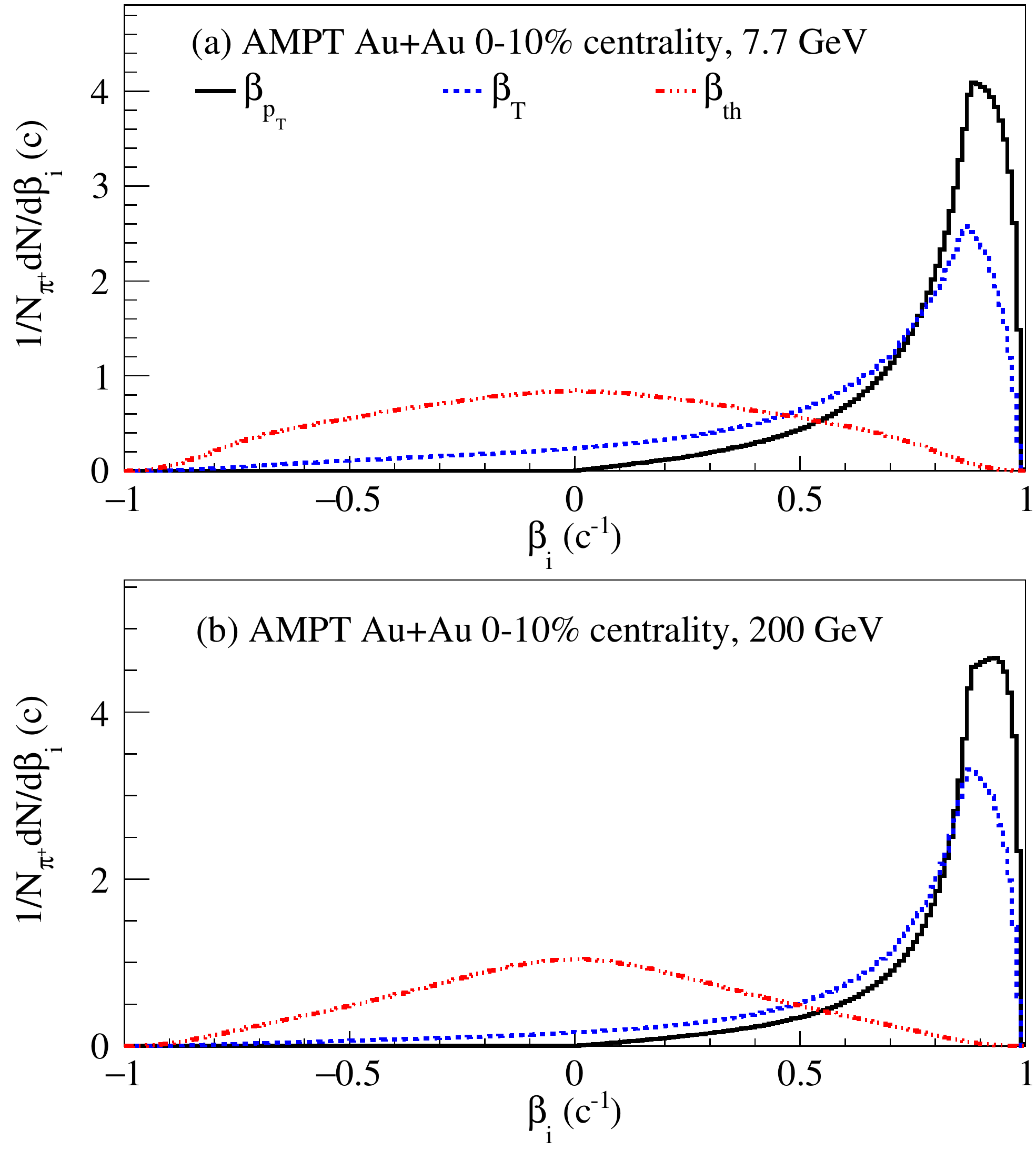}
	\caption{The $\beta_{p_T}$, $\beta_{T}$ and $\beta_{th}$ distribution for $\pi^+$ system at $\sqrt{s_{NN}}$ = 7.7 and 200 GeV in 0-10\% central Au + Au collisions by using the String Melting AMPT model.}  
	\label{fig:beta_T}
\end{figure}

(4) The total energy of system is mainly dominated by
transverse momentum with $\pi^+$ at mid-rapidity, which
can be concluded from the transverse velocity distribution for $\pi ^+$ as shown in Fig.~\ref{fig:beta_T}, most $\pi^+$ are distributed in the region over $\beta_{p_T} > 0.8c$. So $\pi^+$'s energy can be written as a function of $p_T$. Considering the radial flow, the total energy can be written as $E = E_{th} + E_{R}$ where $E_{th}$  is internal energy from thermal motion and $E_{R}$ is energy from collective radial flow \cite{Cv_models_mean_pT_Basu}. In thermodynamical viewpoint, the heat capacity should only be correlated to $E_{th}$. { We can take a quick glance for the radial collective motion and thermal motion of the system via giving distributions of radial flow $\beta_T$ and thermal velocity $\beta_{th}$ = $\frac{ \boldsymbol{\beta}_{\boldsymbol{p_T}}\cdot \boldsymbol{\rho_T}}{|\boldsymbol{\rho_T}|}$ where $\boldsymbol{\rho_T}$ and $\boldsymbol{\rho}$ constitute a right-handed orthogonal system. As shown in Fig.~\ref{fig:beta_T}, the pions are moving thermally in a  expanding fireball with a collective radial velocity.} To avoid complex differential term of $E_{R}$, we firstly use total energy $E$ to derive a formula  of an effective heat capacity. The following calculation will show us that, though the derived effective heat capacity includes contribution from $E_{R}$, the radial flow still can not dominate the energy dependence and fluctuation properties of heat capacity. This fact ensures our calculation and discussion on new derived heat capacity reasonable.

The total energy of a $\pi^+$ system can be written as~\cite{AMPT2005,AMPT_temperature_parton_Lin}
\begin{eqnarray}
\label{eq:E_total_def}
\begin{aligned}
E = \sum_{i=1}^{N}E_i = \sum_{i=1}^{N}\left(p_{Ti}^2 + p_{zi}^2 + m_0^2\right)^\frac{1}{2}.
\end{aligned}
\end{eqnarray}
By calculating the variance and correlation coefficient of $T_{kin}$ and radial flow, we can prove that in the following derivation of $C_{v}$, the $E_{th}$ can be replaced by total energy $E$ in Eq.~(\ref{eq:E_total_def}) reasonably.

 Here we discuss the contribution in energy from radial flow. It should be noted that we did not remove $f(\langle\beta_{T}\rangle)$ from $E$ directly because our $\beta_{T}$ is not exactly same as the one in the Blast-Wave Model in energy composition (though in temperature calculation our $\beta_{T}$ approximately equals to $\beta_{T}$ from the Blast-Wave model \cite{pT_spectrum_mean_pT_Tkin_beta_Adam,Blast_Wave_pT_Retiere}). Thus we can not simply write the energy contributed by radial flow as form of non-relativistic kinetic energy~\cite{Blast_Wave_pT_Retiere,pT_beta_Tkin_Abelev,pT_beta_Tkin_Adam}.  

\begin{figure}[htbp]
	\centering
	\includegraphics[angle=0,scale=0.45]{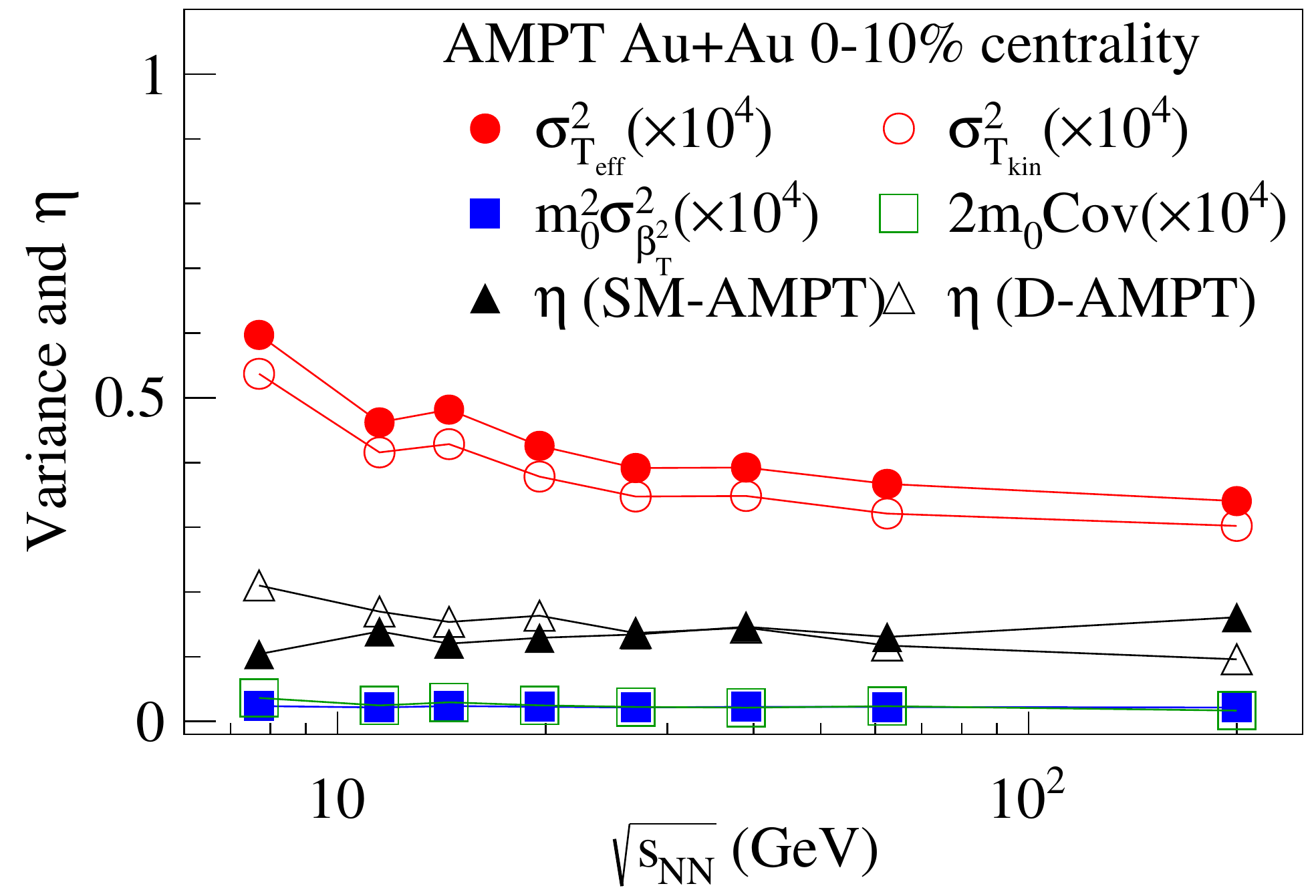}
	\caption{The variance of each term in temperature from the String Melting AMPT model, and the correlation coefficient between $T_{kin}$ and $\beta_{T}^2$ in different versions.  }  
	\label{fig:var_terms}
\end{figure}

In Eq.~(\ref{Teff_Tkin_m0}), the variance of $T_{eff}$ can be written as,
\begin{eqnarray}
\begin{aligned}
\sigma^2_{T_{eff}}\approx \sigma^2_{T_{kin}} + m_0^2\sigma^2_{\langle\beta_{T}\rangle^2} + 2m_0\mathrm{Cov}\left(T_{kin},\langle\beta_{T}\rangle^2\right).
\end{aligned}
\label{eq:variance_Teff}
\end{eqnarray}
 Fig.~\ref{fig:var_terms} shows the variance of $T_{eff}$ and each term in the right side of Eq.~(\ref{eq:variance_Teff}). We see that the variance from $T_{kin}$ dominates the one for $T_{eff}$ while the one from $\langle\beta_{T}\rangle^2$ is less with one order of magnitude, thus the variance   of $\langle\beta_{T}\rangle$ could be negligible for properties of temperature.
In order to clarify that the radial flow term is independent from temperature term in our derivation and calculation, we investigate 
the linear correlation coefficient $\eta = \mathrm{Cov}(T_{kin},\langle\beta_{T}\rangle^2) / (\sigma_{T_{kin}}\sigma_{\langle\beta_{T}\rangle^2})$ which is also presented in Fig.~\ref{fig:var_terms}, here $\eta \leq 0.2$ means that they are linear independent. In addition {the derivative}  of $T_{eff}$ over $T_{kin}$ is approximately a constant,
then we can use $T_{eff}$ instead of $T_{kin}$ for differential term from radial flow.

According to our assumptions, a $\pi ^+ $ system evolves along a continuous path on $\sqrt{S_{NN}}$. Each system with its evolution path corresponds to their own kinetic parameters like $E$, $T$ and $\langle p_{T}\rangle$,  when $\sqrt{s_{NN}}$ changes, the total energy $E$ of $\pi^+$ system changes to $E+\Delta E$ on the continuous path. Similarly, when  the $p_{T}$ spectrum changes, resulting $p_{T}$ of each $\pi^+$ in one system changes to $p_{T} + \Delta{p_{T}}$. Here we see the advantage of the merged events, which ensured similar multiplicity when $E$ changes. That means for each $\pi^+$ on the continuous evolution path we can find a closest new  $\pi^+$ in phase space with $p_{T} + \Delta{p_{T}}$. In this sense, $E$ and $p_{T}$ for each $\pi^+$ can be seen as a continuous function of system temperature.

{\par Under the above four assumptions, especially considering the fact that the energy only depends on $p_T$ in mid-rapidity, by using Eq.~(\ref{eq:Cv_def}) and Eq.~(\ref{eq:E_total_def}) we get}
\begin{eqnarray}
\label{eq:derivation_Cv}
\begin{aligned}
\frac{\partial E}{\partial T}\vert_{E\left(T\right)}&=\sum_{i}^{N}\frac{\partial}{\partial T}\left(p_{Ti}^2+p_{zi}^2+m_0^2\right)^\frac{1}{2}\\
&=\sum_{i}^{N}\frac{\partial E_i}{\partial p_{Ti}}\frac{\partial p_{Ti}}{\partial T}
=\sum_{i}^{N}\frac{1}{2}\frac{2p_{Ti}}{E_i}\frac{\partial p_{Ti}}{\partial T}\\
&=\sum_{i}^{N}\left(\beta_{p_{Ti}}\frac{\partial p_{Ti}}{\partial T}\right)\vert_{E\left(T\right)}.
\end{aligned}
\end{eqnarray}

The sum can be converted to the integral pattern with help of Riemann Sum. Let $f\left(x_i\right)=F\left(x_i\right)/{\Delta x_i}$, for equidistant division, the Riemann integral can be written as
\begin{eqnarray}
\begin{aligned}
\int_{a}^{b}f\left(x\right)g\left(x\right)dx\approx\sum_{i}^{\left(b-a\right)/{\Delta x}}\frac{F\left(x_i\right)}{\Delta x}g\left(x_i\right)\Delta x .
\end{aligned}
\label{eq:11}
\end{eqnarray} 
It should be noted here $N_x = \frac{\left(b-a\right)}{\Delta x}$ stands for the number of division in Riemann integral which is different from multiplicity $N_{multi}$, but we can approximately let $N_{x} = N_{multi}$, so that the definition of $C_v$ expressed by sum can be related to integral over temperature. 
As mentioned in Ref.~\cite{Cv_pT_HM_QGM_Sa} and Ref.~\cite{Cv_models_mean_pT_Basu}, merged events ensured enough high multiplicity and then small error for fitting. 

Using Eq.~(\ref{eq:11}), Eq.~(\ref{eq:derivation_Cv}) can be expressed as follows,
\begin{eqnarray}
\begin{aligned}
&\sum_{i}^{N_{multi}}\beta_{p_{Ti}}\frac{\partial p_{Ti}}{\partial T}\vert_T
=\sum_{i}^{\left(b-a\right)/{\Delta x}}\frac{\beta_{p_{Ti}}}{\Delta T}\frac{\partial p_{Ti}}{\partial T}\Delta T\\
&\approx \int_{a}^{b}\frac{\beta_{p_{Ti}}}{dT}\frac{\partial p_{Ti}}{\partial T}dT
=\left.\left(\frac{\beta_{p_{Ti}}}{dT}\frac{\partial p_{Ti}}{\partial T}\right)\right|_{T={T_0}}\left(b-a\right)\\
&\left.\approx \tilde{\beta_{p_{T}}}\frac{\partial {\tilde{p_{Ti}}}}{\partial T}\right|_{T={T_0}}\frac{\left(b-a\right)}{\Delta T}
= N_{multi}\left.\left(\tilde{\beta_{p_{T}}}\frac{\partial {\tilde{p_{Ti}}}}{\partial T}\right)\right|_{T={T_0}}.
\end{aligned}
\end{eqnarray}
\par Here $T_{0}$ means the temperature in characteristic event $'\xi'$, corresponding to the $'\xi'$ noted in Mean Value Theorem of Integrals. 

The Mean Value Theorem of Integrals says:
Let $f:\left[a,b\right]\rightarrow\mathbb{R}$ be a continuous function, $g:\left[a,b\right]\rightarrow\mathbb{R}$ be integrable and constant-sign, then there exists $\xi \in\left(a,b\right)$ so that $\int_{a}^{b}f\left(x\right)g\left(x\right)dx = f\left(\xi\right)\int_{a}^{b}g\left(x\right)dx$.

According to  Mean Value Theorem of Integrals , we can prove the existence of  $\tilde{\beta_{p_{T}}}$ and $\tilde{p_{Ti}}$, these two parameters reveal the evolution property and energy dependence of system.
 Hence the derivation gave the specific heat expressed by kinetic quantities:
\begin{eqnarray}
\label{eq:Cv_xi}
\begin{aligned}
C_v^{T_0} = \frac{C}{N_{multi}}\approx \tilde{\beta_{p_{T}}}\left({T_0}\right)\left.\frac{\partial \tilde{p_T}}{\partial T}\right|_{T_0}.
\end{aligned}
\end{eqnarray}

 In Eq.~(\ref{eq:Cv_xi}) and Eq.~(\ref{eq:derivation_Cv}), the heat capacity for ensemble can be expressed as a combination of kinetic measurements in a characteristic event whose $T = {T_0}$.  

The above equation gave a formula for calculating specific heat approximately by transverse velocity, transverse momentum and temperature in a unique event $'\xi'$, i.e. by a derivative of transverse momentum to temperature in a unique event $'\xi'$. It should be noted here that $C_v^{T_{0}}$ apparently includes radial flow term by $\beta_T$, but  $\beta_{T}$ decreases at low $\sqrt{s_{NN}}$ ~\cite{pT_spectrum_mean_pT_Tkin_beta_Adam} and $f(\langle \beta_{T} \rangle)$ is smaller than $T_{kin}$, the dominant contribution is actually from thermal motion. That is why in Eq.~(\ref{eq:Cv_def}) we can use $E$ instead of $E_{th}$.

However, the first mean value theorem for integrals can only provide the existence of $'T_{0}'$, it can not find the accurate value to calculate the $C_v^{T_{0}}$. Therefore the present study needs to further use the mean value of kinetic quantities instead of unique one to give accessible results that can be measured and computed from data:
\begin{eqnarray}
\begin{aligned}
\label{eq:Cv_xi*}
{C_v^{*}}=\overline{\beta_{p_{T}}}\left.\frac{\partial \overline{p_T} }{\partial T}\right|_{T=\overline{T_{eff}}} .
\end{aligned}
\end{eqnarray}

We should emphasize that the definition in Eq.~(\ref{eq:Cv_xi*}) is a compromise for the real calculation of  $C_v^{T_{0}}$. Since $\xi$ in the Mean Value Theorem for Integrals is unsolvable,  we use the mean value of $\beta_{p_{T}}$ and $T_{eff}$ instead of values in single event 
to make sure $C_v^{T_{0}}$ can be approximately calculated in practice.

 According to the definition in Eq.~(\ref{eq:Cv_def}), we should use $T_{kin}$ as $T$ in calculation of $\overline{p_T}$ in Eq.~(\ref{eq:Cv_xi*}). However,  this formula involving radial flow can not give identical results of $\frac{\partial \overline{p_T}}{\partial T_{kin}}$.  Fortunately, as demonstrated in Fig.~\ref{fig:var_terms}, it can be calculated by the AMPT result that derivatives for $T_{kin}$ and $T_{eff}$ have strongly linear dependence, we can directly use $T_{eff}$ instead of $T_{kin}$ in $\frac{\partial \overline{p_T}}{\partial T}$. Now we can conveniently calculate $\frac{\partial \overline{p_T}}{\partial T}|_{T=\overline{T_{eff}}}$ based on  $p_T$ spectrum~\cite{pT_spectrum_mean_pT_Tkin_beta_Adam,mean_pT_1,mean_pT_Teff_Gardim,dpT_dT}, 
\begin{small}
\begin{eqnarray}
\begin{aligned}
\label{eq:<p_T>}
&\overline{p_T}=\frac{\int_{p_{min}}^{p_{max}}p_T^2F\left(p_T\right)dp_T}{\int_{p_{min}}^{p_{max}}p_TF\left(p_T\right)dp_T}
	= 2T_{eff} + \\ 
	&\frac{p_{min}^2e^{-p_{min}/T_{eff}}-p_{max}^2e^{-p_{max}/T_{eff}}}{\left(p_{min}+T_{eff}\right)e^{-p_{min}/T_{eff}}-\left(p_{max}+T_{eff}\right)e^{-p_{max}/T_{eff}}} .
\end{aligned}
\end{eqnarray}
\end{small}

Considering that the mean value and unique value are obviously different, which directly connected to our explanation on the statistical parameters, we will discuss the topic in the next section.

\section{Results of  $C_v^{*}$ and its high order moments}
\label{sec:results}

\begin{figure}[htbp]
	\centering
	\includegraphics[angle=0,width=9cm,height=9cm]{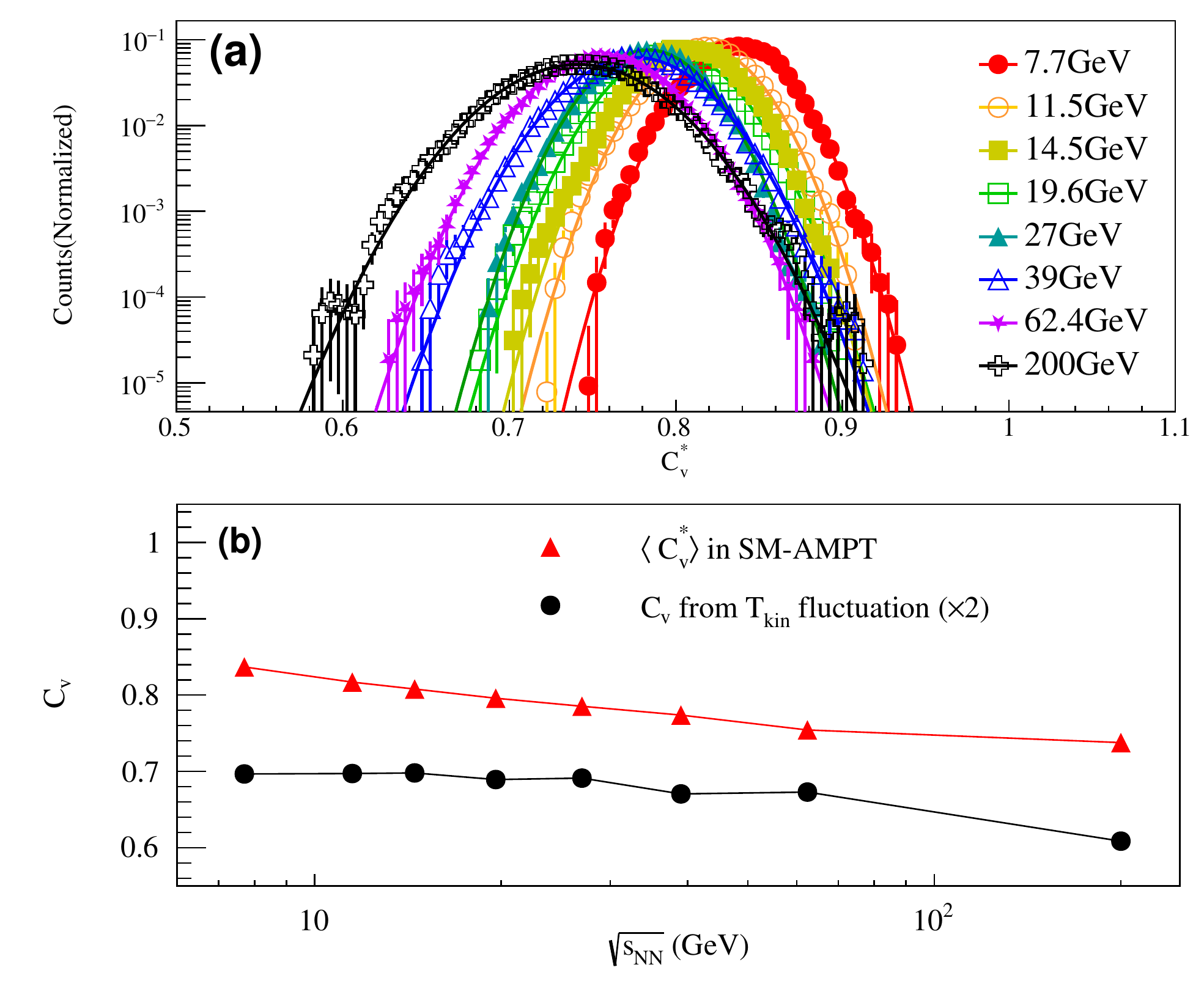}
	\caption{(a) The event-by-event  $C_v^{*}$  distribution for $\pi^+$ system in $0-10\%$ central Au+Au collisions at  $\sqrt{s_{NN}}$ = 7.7, 11.5, 14.5, 19.6, 27, 39, 62.4, and 200 GeV, respectively,  by using the String Melting AMPT. The lines are fitted by Gaussian distribution; 
	(b) The extracted $\langle C_v^{*}\rangle$ from  the String Melting AMPT model calculation in red line, compared to the $C_v$ from $T_{kin}$ fluctuation which is represented by the black line, corresponding to the black star marker in Fig.~\ref{fig:Cv_models}. Note that the $C_v$  value from $T_{kin}$ fluctuation are scaled by twice. }  
	\label{fig:Cv_xi_distribution}
\end{figure}

Using the derived formula~(\ref{eq:Cv_xi*}) and~(\ref{eq:<p_T>}), $C_v^{*}$ can be calculated with data of $\pi ^+$ system in Au + Au collisions at different $\sqrt{s_{NN}}$ simulated by the AMPT.  The event-by-event distribution of $C_v^{*}$ presents the Gaussian-like distribution as shown in Fig.~\ref{fig:Cv_xi_distribution}, from which  
the mean values ($\langle C_v^{*}\rangle$) at each $\sqrt{s_{NN}}$ can be obtained. 

Figure~\ref{fig:Cv_xi_distribution}(b) shows $\left\langle C_v^{*}\right\rangle$ from the AMPT calculations, which displays a slightly decreasing trend with the increasing of $\sqrt{s_{NN}}$. Compared to the results in Ref.~\cite{Cv_pT_HM_QGM_Sa}, the energy dependence of $\left\langle C_v^{*}\right\rangle$ is similar to the results of QGM, HM and HM via QGM in Fig.~\ref{fig:Cv_models}, even though the values of $C_{V}$ and $\left\langle C_v^{*}\right\rangle$ are slightly different. In Ref.~\cite{Cv_pT_HM_QGM_Sa}, the specific heat from QGM is much different from results in pQCD thermodynamic method and approximate pure gauge theory, which should be studied further to confirm the discrepancy contributed by classical and quantum statistics, the latter two theoretical calculations both gave larger $C_V (10\sim15)$ at low $\sqrt{s_{NN}}$ than other  models shown in Fig.~\ref{fig:Cv_models}. So they suggested using { $p_{T}$ distribution of $\pi^+$ in RHIC single events} to extract $C_{V}$ according to the same statistical method as Ref.~\cite{Cv_pT_HM_QGM_Sa}. For our calculation of $C_v^{*}$, we suggest extracting transverse velocity and effective temperature from $p_{T}$ spectrum. 
But as far as the uncertainty for present experimental data, if one considers to extract parameters for $C_v^{*}$ which we discussed above, we suggest to improve experimental precision to ensure smaller uncertainty for fitting parameters. And if possible, the direct particle identification, transverse velocity measurement and $p_T$ spectra  for merged events of charged $\pi$ system in experiments are also recommended, so that we can check our methods and fit  experimental results as accurate as possible.

\par
\begin{figure}[htbp]
	\centering
	\includegraphics[angle=0,scale=0.45]{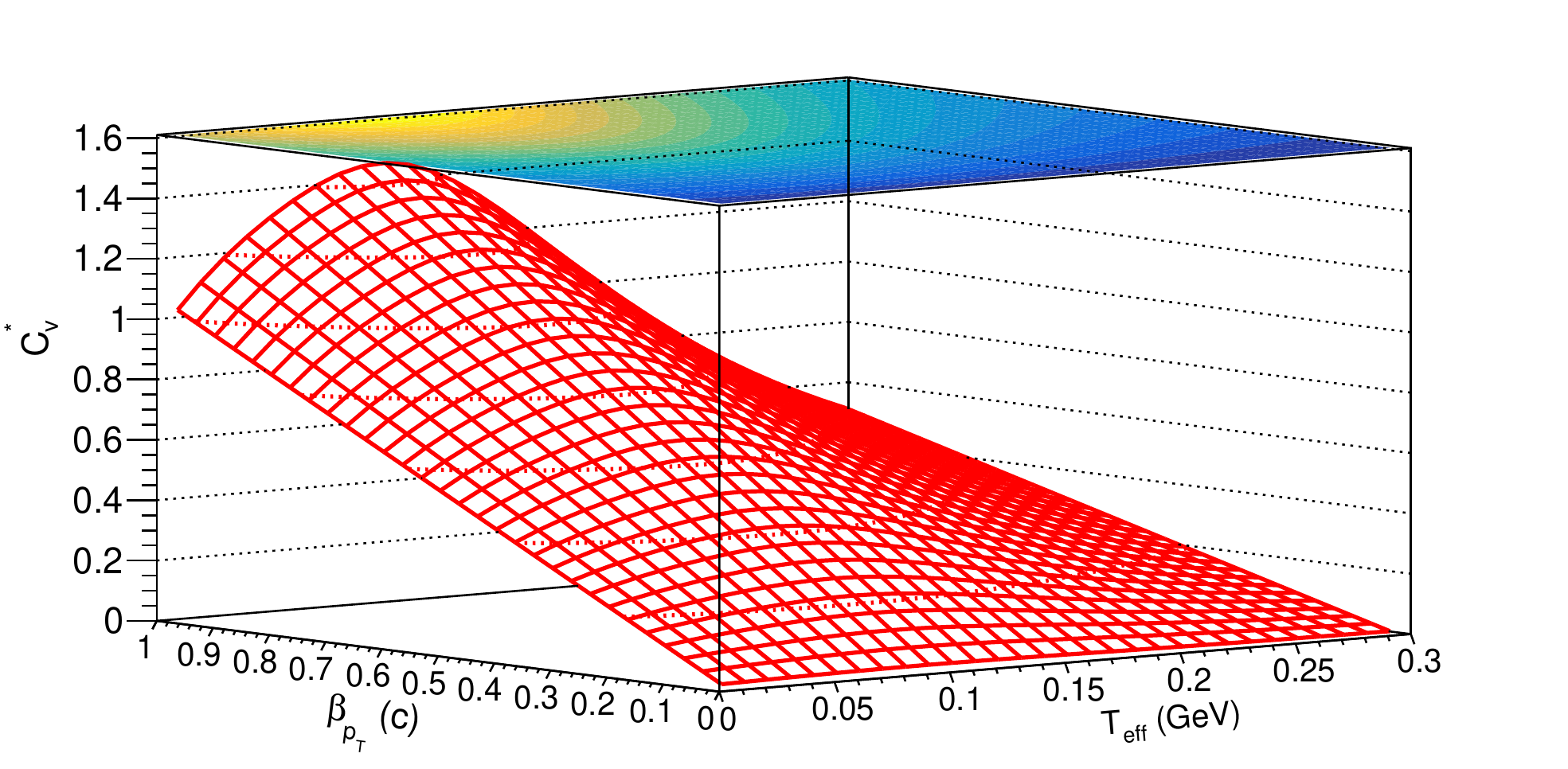}
	\caption{Theoretical $C_v^{*}$ distribution as a two-dimensional function of $\beta_{p_{T}}$ and $T_{eff}$.}  
	\label{fig:Cv_xi_funcplot}
\end{figure}

 To clearly illustrate the sensitivity of $C_v^{*}$ to $\beta_{p_T}$ and $T_{eff}$, we plot $C_v^{*}$ as functions of {$\beta_{p_T}$ and $T_{eff}$} by using Eqs. (\ref{eq:Cv_xi*}) and (\ref{eq:<p_T>}) in Fig.~\ref{fig:Cv_xi_funcplot}. An inflection point exists at around $T_{eff} = 100$ MeV, which depends on the fitting limits of $p_{T}$ spectrum. The combination of linear dependence of {$\beta_{p_T}$} and non-monotonic dependence of $T_{eff}$ results in that $C_v^{*}$ is sensitive to the range of the fitting parameters. Along an identified curve of $C_v^{*}$ in Fig.~\ref{fig:Cv_xi_funcplot}, a shifting fitting result may cause different trend of  energy dependence of $C_v^{*}$, which has to be studied further, especially near the QCD critical point.

 In Fig.~\ref{fig:Cv_xi_distribution}(b) 
 the mean values of $C_v^{*}$ are smaller than those extracted from the HRG model, HM model as well as  QGM model~\cite{Cv_pT_HM_QGM_Sa}. In the first mean value theorem for integrals, mean value deviates the ideal $C_v^{T_{0}}$. That means only if one can accurately find the evolution curve of a characteristic event corresponding to temperature $T_{0}$, {the real $C_v^{T_{0}}$ can agree with results extracted by temperature fluctuation from the STAR's data especially at low $\sqrt{s_{NN}}$ ~\cite{Cv_models_mean_pT_Basu}.}
It is interesting to see that $\langle C_v^{*}\rangle$ remains the rising trend with the decreasing of $\sqrt{s_{NN}}$, which suggests that in the event-by-event distribution of $C_{v}^{*}$ we can still find special properties similar to  sharp decreasing trend with increasing of colliding energy~\cite{HRG_model_Mait}. 

As a comparison, the behaviour of $C_v^{*}$ with parton process or only with hadron process are different. Though the rising trend of average $C_v^{*}$ is depressed, the events with parton process  provides higher $C_v^{*}$ similar to the cases of pQCD and other models~\cite{Cv_models_mean_pT_Basu,Cv_pT_HM_QGM_Sa}. That means that more events with higher $C_v^{*}$ emerge at low $\sqrt{s_{NN}}$ in the SM-AMPT case. Further, 
we can investigate the fluctuation behaviour of $C_v^{*}$ by event-by-event distribution as we did in temperature distribution. And indeed we can see nontrivial fluctuation behaviour of $C_v^{*}$ when taking parton interaction into account.

\begin{figure}[htbp]
	\centering
	\includegraphics[angle=0,width=9cm,height=9cm]{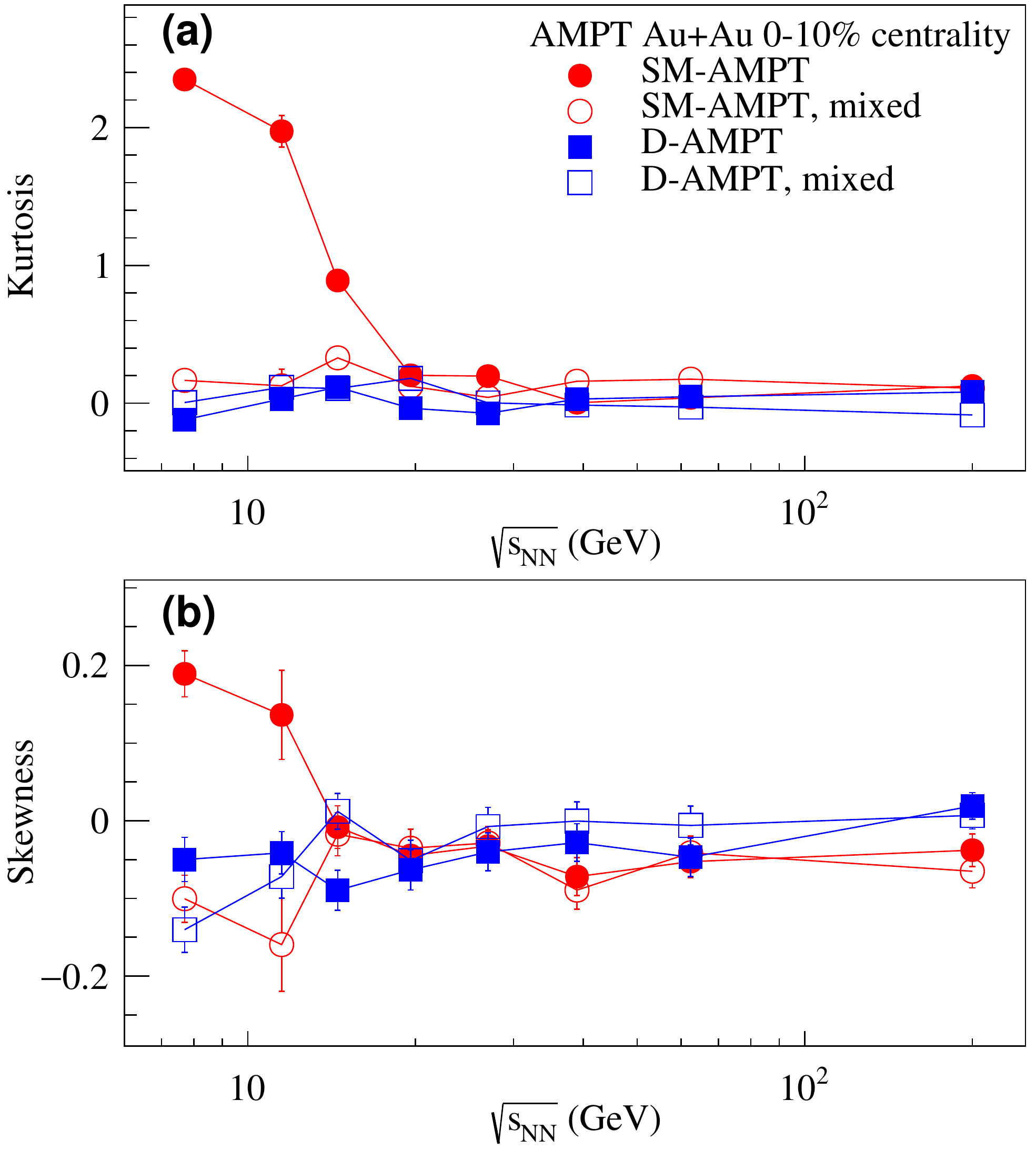}
	\caption{The energy dependence of kurtosis (a) and  skewness (b) of $C_v^{*}$ for $\pi^+$ system from the String Melting AMPT (red circle) is  compared to results from the Default AMPT version (blue square) as well as random mixed statistical fluctuations (red open circle and blue open square).}
	\label{fig:skewness}
\end{figure}

Based on the event-by-event distribution of $C_v^{*}$, skewness and kurtosis of $C_v^{*}$ can be obtained in the AMPT model with the String Melting and Default version, respectively. 
The skewness and kurtosis are defined as $ \frac{\mu_{3}}{\sigma^3}$ and $\frac{\mu_{4}}{\sigma^4}-3$, respectively. Here $\mu_n = \langle ( X- \langle X \rangle )^n \rangle$ and $\sigma = \sqrt{\frac{\Sigma (X_i-\langle X\rangle)^2}{N}}$ where the $X$ represents  $C_v^{*}$. The skewness reflects the deviation of $C_v^{*}$ distribution from the Gaussian distribution, and the kurtosis describes how close the events distribute to expectancy (a standard Gaussian distribution has zero kurtosis).
At RHIC, skewness and kurtosis analysis have been successfully applied to net-proton fluctuation to explore possible QCD critical point \cite{Ste,STAR_fl1,STAR_fl2,STAR2021,Luo2017}. Here we use this kind of high-order moments to specific heat analysis.
 Fig.~\ref{fig:skewness}(a) and (b), respectively, showed the skewness 
 and kurtosis 
 as a function of $\sqrt{s_{NN}}$, which  exhibits obviously different at lower $\sqrt{s_{NN}}$. In the String Melting version, the skewness and kurtosis all showed a sharp enhanced structure with the decreasing of $\sqrt{s_{NN}}$ below 20 GeV. Meanwhile in the Default version, the skewness and kurtosis are both close to zero and present an independence of $\sqrt{s_{NN}}$. Again, the parton interaction demonstrates its important role for  $C_v^{*}$ fluctuation. 
 
 Before discussing the statistical properties obtained from $C_v^{*}$, we need to take care  statistical fluctuation. As  Ref.~\cite{Cv_models_mean_pT_Basu} shows, the contribution of statistical fluctuation in $T_{eff}$ distribution can be written as $ \left(\Delta T_{eff}\right)^2 = \left(\Delta T^{dyn}_{eff}\right)^2 + \left(\Delta T^{sta}_{eff}\right)^2 $
\label{eq:stat fluc}. 
In Ref.~\cite{Cv_models_mean_pT_Basu}, the statistical fluctuation was extracted by randomly mixing  data from experiments, and then the width of the randomly mixed temperature distribution was removed. 
Here we randomly mixed each particle from a single event while kept the multiplicity in each mixed event  is close to the one in our merged events at about 1000, then we extracted skewness and kurtosis from the mixed events to compare with the results from two versions of AMPT.

In Figure~\ref{fig:skewness}(a) and (b),  
we can clearly find the results from the randomly mixed events and the Default version AMPT are close to each other, which means the energy dependence of skewness and kurtosis in the Default version without partonic process are mainly contributed by statistical fluctuation. Meanwhile, at lower $\sqrt{s_{NN}}$ the String Melting version showed clearly an enhancement structure far from both the Default version and random mixed one. This comparison demonstrates that the unique behavior of 3rd and 4th order moment from the String Melting version at lower $\sqrt{s_{NN}}$ are mainly from the dynamical fluctuation, which reflected that the special properties of thermodynamic  quantities indeed survived throughout partonic and hadronization processes -- even partly. Besides we can also compare the Default result and the random mixed one,  and see only slight difference. The slight difference shows us that the self-correlation of merged events can slightly suppress kurtosis and skewness at low $\sqrt{s_{NN}}$. 
  
The difference between two versions, namely with or without partonic interaction process, shows different thermal properties. When $\sqrt{s_{NN}}$ decreases, skewness of $C_v^{*}$ from the String Melting version increases rapidly to a high positive value. That means, compared to the default version, more events with higher $C_v^{*}$ distributed on the right side of mean value  as illustrated in Fig.~\ref{fig:Cv_xi_distribution}~(a). As mentioned in this section, the real $C_v^{T_{0}}$ obtained from $'\xi'$ was depressed by averaging over the events, but the positive skewness indicated that parton phase in the String Melting version contributes higher value of $C_v^{T_{0}}$ than that in the Default AMPT version. The higher value indicates that, in the String Melting version of AMPT, the heat capacity of $\pi^+$ system behaves more similarly to prediction in other models or theories \cite{Cv_models_mean_pT_Basu,Cv_pT_HM_QGM_Sa} than the Default version. Here  the positive skewness is explained as a signal for formation of $\pi^+$ system closer to the real $C_v^{T_{0}}$ state, instead of explained as overall enhancement of $C_v^{T_{0}}$ in $\pi^+$ system. Because if it is the latter case, one should find the event-by-event distribution of   $C_v^{*}$ will move totally into a higher range and still follow the standard Gaussian distribution.

 {For the distribution of kurtosis of $C_v^{*}$,} it is close to zero at high $\sqrt{s_{NN}}$ and rose rapidly to large
 positive values below 20 GeV  in the String Melting version. Meanwhile we can see that the kurtosis from the Default version keeps close to zero at both high and low $\sqrt{s_{NN}}$. The low kurtosis in the Default version indicates that if we choose AMPT with hadron gas phase instead of parton phase, the $C_v^{*}$ distribution shows similarity to the standard Gaussian distribution. The sharp enhancement in the SM version shows that the $\pi^+$ system formed after parton phase drives events to a distribution with much smaller deviation ($\sigma$) than the one formed from hadron gas. Comparing to Eq.~(\ref{eq:C_fluctuation}) it can be found that though $\sigma$ for $C_v^{*}$ and $T_{kin}$ or $T_{eff}$ are different, they both reflect the significant change of event-by-event fluctuation of the system. As in  results from other models or data the enhancement structure of $C_v$ at lower $\sqrt{s_{NN}}$ demonstrates that the fluctuation of temperature extends to system size of all events in ensemble, and the enhancement structure of $C_v^{*}$'s kurtosis indicates that the fluctuation of $C_v^{*}$ behaves similarly~\cite{temperature_fluctuation_thermal_A,temperature_fluctuation_Ayala1,temperature_fluctuation_Stodolsky}. The results also illustrate that the properties of charged particles formed via parton phase are significantly different from those experienced no partonic process. We explain the difference as a result of partonic process, the introduction of parton provides different degrees of freedom in hot dense matter. However, introducing parton only is not the same as the contribution from critical phase transition, that is why we can not extract the divergence of $C_V$ in the frame. But fortunately, the significant rising trend of skewness and kurtosis in String Melting version at lower $\sqrt{s_{NN}}$ shows us that the effect of partonic process can survive hadronization. And further skewness and kurtosis imply us that change for the degrees of freedom resulted from parton phase can be measured in form of high-order moments of $C_v^{*}$. These observables show clear energy dependence than $C_V$ from temperature fluctuation since we can obtain $C_v^{*}$ event-by-event distribution as many as we need. For  temperature fluctuation, however, we can only extract one $C_V$. In this context, one can clearly reveal the effect from parton process by investigating $C_v^{*}$ and its high-order moments. In one word, the energy dependence of $C_v^{*}$ together with its skewness and kurtosis could be taken as a potential  and verifiable signal to investigate the nature around the critical point of hot dense matter created in heavy-ion collisions. 
 
\section{Summary}
\label{sec:summary}

In this work, we studied the energy dependence of specific heat in a framework of AMPT model. Specific heat  was obtained via event-by-event temperature  fluctuation for $\pi^+$ system from the AMPT model and compared with those from the STAR's data \cite{Cv_models_mean_pT_Basu} and other models \cite{Cv_pT_HM_QGM_Sa}, the difference  among various models and data was analyzed and the  reasons were presented. Based on a few assumptions, we derived a new quantity, namely $C_v^{*}$, to describe specific heat from kinetic quantities of  characteristic event.  By using the AMPT model,  
the $\sqrt{s_{NN}}$ dependence of $\langle C_v^{*}\rangle$ was presented.
This work found that the $\langle C_v^{*}\rangle$ behaves closer to results from the HRG and STAR's data than $\langle C_v\rangle$ from temperature fluctuation in the AMPT frame. But at low $\sqrt{s_{NN}}$, comparing to results from the HRG or STAR's data, our results for $\langle C_v^{*}\rangle$ still seem too small to show a clearer energy dependence. The too small values of $\langle C_v^{*}\rangle$
 are not the ideal results but within the expectation. 
 
In terms of definition in Eq.~\ref{eq:Cv_xi}, our original goal of the  extraction of heat capacity should be the $\xi$ in the Mean Value Theorem for Integrals, while we can only use the mean values of $\beta_{p_T}$ and $T_{eff}$ as alternative substitutes, because the real $\xi$ is unsolvable. As a result of using the mean value,  $\langle C_v^{*}\rangle$ shows a visible energy dependent trend. In other words, if one can find the event $\xi$ and use the parameters in this event $\xi$ to calculate $C_v^{T_{0}}$, we would see the same energy dependence of system as the results in experiments.
 
We further gave the event-by-event distribution of $C_v^{*}$ and then obtained skewness and kurtosis of the $C_v^{*}$ distribution. To ensure the calculation reasonable, it was additionally compared for the variance of different terms and linear correlation coefficient. We compared parameters like skewness and kurtosis in two different versions of AMPT.

The results demonstrate that the energy dependences of kewness from the String Melting version  show clear difference from the Default version, and similar energy dependence can be seen in kurtosis. The above phenomenon indicates that partonic process drives more $\pi^+$ events emerge in the higher specific heat state. And from kurtosis it can be seen that the behavior of fluctuation and deviation of $C_v^{*}$ in the String Melting version showed clearer energy dependence, which is similar to the behavior of fluctuation of temperature described by $C_v$ from results of Refs.~\cite{temperature_fluctuation_Basu,Cv_mean_pT_Pythia_ALICE_Deb,Cv_models_mean_pT_Basu,Cv_pT_HM_QGM_Sa}.
  
In  future, we  expect more experimental and theoretical  investigation by  using  heat capacity to search for the QCD phase transition of the hot dense matter created in heavy-ion collisions.

\begin{acknowledgements}
This work was supported in part by the National Natural Science Foundation of China under contract Nos.  11890710, 11890714, 11875066, 11925502, 11961141003 and 12147101, the Strategic Priority Research Program of CAS under Grant No. XDB34000000, National Key R\&D Program of China under Grant No. 2018YFE0104600 and 2016YFE0100900, and by Guangdong Major Project of Basic and Applied Basic Research No. 2020B0301030008.

\end{acknowledgements}

	\end{CJK*}	

\bibliography{temperaturefluctuation}

\end{document}